\documentclass{he_symp}
\usepackage{psfig}
\usepackage{graphicx}  

\def\ergs{{\rm erg}~{\rm s}^{-1} }
\def\ergscm2{~{\rm erg}~{\rm s}^{-1}~{\rm cm}^{-2} }
\def\gcm2{~{\rm g}~{\rm cm}^{2} }
\def\GeV{\rm{GeV}}
\def\MeV{\rm{MeV}}
\def\keV{\rm{keV}}

\def\G{\rm{G}}

\def\K{\rm{K}}
\def\s{\rm{s}}
\def\cm{\rm{cm}}
\def\Lg{$L_\gamma$}

\def\edot{$L_{\rm sd}$}
\def\lambdabar{\mathrel{\lower 1pt\hbox{$\mathchar'26$}\mkern-9mu
        \hbox{$\lambda$}}}
\def \bs {B_{\rm s}}
\def \bpc {B_{\rm pc}}
\def \bcr {B_{\rm crit}}

\def \rns {R_{\rm s}}

\def \eps {{\cal \scriptstyle E}}
\def \epr {\epsilon^\prime}
\def \eb  {\epsilon_{\scriptscriptstyle B}}
\def \st  {\sigma_{\scriptscriptstyle T}}
\def \skn  {\sigma_{\scriptscriptstyle KN}}

\def \sss {\scriptscriptstyle}

\newbox\grsign \setbox\grsign=\hbox{$>$} \newdimen\grdimen \grdimen=\ht\grsign
\newbox\simlessbox \newbox\simgreatbox \newbox\simpropbox
\setbox\simgreatbox=\hbox{\raise.5ex\hbox{$>$}\llap
    {\lower.5ex\hbox{$\sim$}}}\ht1=\grdimen\dp1=0pt
\setbox\simlessbox=\hbox{\raise.5ex\hbox{$<$}\llap
     {\lower.5ex\hbox{$\sim$}}}\ht2=\grdimen\dp2=0pt
\setbox\simpropbox=\hbox{\raise.5ex\hbox{$\propto$}\llap
     {\lower.5ex\hbox{$\sim$}}}\ht2=\grdimen\dp2=0pt
\def\simgreat{\mathrel{\copy\simgreatbox}}
\def\simless{\mathrel{\copy\simlessbox}}

\def\la{\mathrel{\hbox{\rlap{\hbox{\lower4pt\hbox{$\sim$}}}\hbox{$<$}}}}
\def\ga{\mathrel{\hbox{\rlap{\hbox{\lower4pt\hbox{$\sim$}}}\hbox{$>$}}}}
\newcommand{\citep}[1]{(\cite{#1})}

\begin{document}

\title{High-energy radiation from pulsars:\\ a challenge to polar-cap models}
\author{B. Rudak,
        J. Dyks
        \and 
        T. Bulik
}
 
\institute{Nicolaus Copernicus Astronomical Center, Warsaw - Toru{\' n}, Poland}

\maketitle    

\begin{abstract} 
Spectacular results of pulsar observations in X-ray and gamma-ray domains
gave a new boost to
theoretical models of pulsar magnetospheric activity. 
A challenging aspect of 
these efforts is that 
lightcurves and broadband energy spectra of the brightest HE sources
exhibit unexpected richness of features, 
making each object almost a unique case to be interpreted with a custom-made model.
This review offers our subjective account of 
how the models of high-energy radiation used
within the framework
of SCLF polar-cap scenarios tackle with these challenges.
We describe major characteristics of all radiative processes
relevant for modeling the observed features.
Then we address successes 
as well as noticeable disadvantages of these models 
upon their confrontation with the available data.
\end{abstract}

\section{Introduction}
A large fraction of galactic X-ray and gamma-ray sources is
associated with neutron stars, in particular with rotation powered pulsars (RPP).
Spectacular results of observational campaigns of RPP with ROSAT, ASCA and CGRO,
and recently with RXTE and CHANDRA induced a new wave of interest in
theoretical models of pulsar magnetospheric activity. 
A challenging aspect of these efforts is that 
lightcurves and broadband energy spectra of the brightest HE sources
exhibit unexpected richness of features making thus each object a unique case pending special treatment.

The pair creation paradigm is a pivotal element in virtually all models of magnetospheric activity of 
radiopulsars (which at the moment constitute a vast majority of known RPPs).
Electron-positron pairs ($e^\pm$-pairs) 
are thought to be responsible for radio emission observed in radiopulsars
which is interpreted as the coherent curvature radiation of $e^\pm$ plasma.
Pairs can be produced in magnetospheric environments  either via photon absorption in a dense field
of soft photons (photon-photon collision) or via photon absorption in a strong magnetic field. 
In either case a supply of hard gamma photons is required in order to fulfill stringent threshold 
conditions for pair creation.
Not all of these photons would be subject
to absorption; many will escape the magnetosphere without any attenuation.
This leads us to expect that all radiopulsars (including low-$B$ millisecond objects) should be
the sources of gamma radiation.
It is up to theoretical models to show whether the expected flux of the radiation 
is interestingly high with respect to the sensitivity of recent and future gamma-ray telescopes.
One should keep in mind, however, that the pair creation paradigm remains
just a paradigm for the time being: recently it was questioned 
by \cite{jessner}, who presented arguments
that for a wide range of surface temperature, magnetic field
strength and spin period, the electrons supplied by neutron star surface via thermionic and field emission
will screen out the accelerating electric field ${\cal E}_\parallel$, limiting it to a residual value.
In consequence, no favourable conditions would exist for magnetic pair production.

This review offers our subjective account of 
how the models of high-energy radiation do perform within the framework
of polar-cap models with space-charge-limited-flow (SCLF).
First, we introduce all ingredients of polar-cap models with SCLF
and we describe major characteristics of all radiative processes
relevant for physical conditions in pulsar magnetospheres.
Then we address successes 
as well as noticeable disadvantages of these models 
upon their confrontation with the data.

\section{Basic parameters}
For the sake of completeness, we recall in this section basic quantities used throughout 
the review.
A neutron star of radius $R_{\rm s}$ and
moment of inertia $I$, spinning with angular velocity $\Omega = 2\pi/P$ which
decreases in time (for whatever the reason) at a rate $\dot \Omega = - 2\pi\, P^{-2}\dot P < 0$,
loses its rotational energy at the rate
\begin{equation}
L_{\rm sd} \simeq 4\times 10^{31}\, I_{45} \dot P_{-15} P^{-3} {\ergs},
\label{b1}
\end{equation}
where $P$ is in seconds, $\dot P_{-15} \equiv \dot P / 10^{-15}$ and $I_{45} \equiv I/10^{45}\gcm2$, 
and \edot~ is the so called spin-down luminosity.
If a magnetic dipole is anchored at the center of a neutron star, with
its magnetic moment $\vec \mu_B$ inclined at angle $\alpha$ to the spin axis $\vec \Omega$,
then the 
strength of the field at the stellar equator is $B_{\rm s} = \mu_B /\rns^3$ (and twice as high at the poles). 
The dipole, rotating in 
a vacuum will lose energy at the rate
\begin{equation}\label{b3}
L_{\rm magn} = {2 \over {3 c^3}}\, B_{\rm s}^2 \sin^2\alpha \, R_{\rm s}^6 \, \Omega^4.
\end{equation}
The quantity $B_{\rm s} \sin\alpha$ can be thus inferred from $P$ and $\dot P$ for
a neutron star with known values of $I$ and $R_{\rm s}$.
Another model, where the dipolar radiation is replaced with a magnetospheric wind of particles \citep{goldreich},
gives a similar result as (\ref{b3}) for an orthogonal case:
\begin{equation}\label{b5}
L_{\rm wind} \approx {1 \over c^3}\, B_{\rm s}^2 \, R_{\rm s}^6 \, \Omega^4
\end{equation}
and therefore is independent of the angle $\alpha$.
Since no observational support exists for $\dot P$ depending on $\sin\alpha$, the
standard approach is to apply the latter model ($L_{\rm sd} = L_{\rm wind}$) to derive the strength of the dipolar 
component of magnetic field
\begin{equation}\label{b6}
B_{12}^2 =  10^{15}\, I_{45}\,R_6^{-6}\, P\, \dot P
\end{equation}
where $B_{12}\equiv B_{\rm s}/10^{12}\G$, and $R_6\equiv R_{\rm s}/10^{6}\cm$.
It is likely that neutron star magnetic fields contain high-order multipoles which
may dominate the dipolar component at the surface level. 
Their relative amplitudes as well as distribution remain, however, unknown.
It will be assumed throughout the paper that the magnetic field is a static-like dipole,
not distorted by
rotational effects or by presence of strong outflowing wind of particles (the latter effect has been
recently invoked to decrease very high values of $B_{\rm s}$ inferred from $P$ and $\dot P$ for two SGRs \citep{kazanas}).
The field is therefore approximated with axisymmetric field lines satisfying
$r \sin^{-2}\theta = R_{\rm dc}$ in polar coordinates $r$ and $\theta$, with the dipole constant $R_{\rm dc}$.
A dipole constant for which a rigid rotation with the angular velocity $\Omega$ reaches the speed-of-light limit 
(it occurs for $R_{\rm dc} = c/\Omega$ and this particular value is denoted as $R_{\rm lc}$) determines the so 
called light cylinder 
of radius $R_{\rm lc}$.
All field lines which cross the light cylinder are then considered as open lines, and their
footpoints on the stellar surface define two 
polar caps of radius $R_{\rm pc} \approx R_{\rm
s}\cdot (R_{\rm s}/R_{\rm lc})^{1/2}$, where the latter factor is the sine function of the magnetic
colatitude 
$\theta_{\rm pc}$ for the outer rim of the polar cap: 
\begin{equation}\label{pc}
\sin \theta_{\rm pc} = (R_{\rm s}/R_{\rm lc})^{1/2}.
\end{equation}

\section{Properties of pulsars in X-rays and gamma-rays}\label{Xgamma}  
For more than 1500 pulsars known to date only about $\sim 40$ positive detections in X-rays and no more than
10 detections in gamma-rays have been achieved (\cite{becker2002}, \cite{kanbach2002}). 
CGRO provided seven high-confidence detections (of {\it Seven Samurai})
with three other cases classified as `likely' detections. The gamma-ray sources
were identified by virtue of flux pulsations with previously known $P$ and $\dot P$. 
Crab and Vela are the only pulsars seen by all three instruments of CGRO.
No trace of pulsed signal in VHE range (300 GeV -- 30 TeV) has been found so far for 
the gamma-ray pulsars  (\cite{sako}, \cite{weekes}). 
However, strong steady VHE emission is associated with 3 out of 10 gamma-ray pulsars.
Two plerionic sources of the steady VHE radiation -- The Crab Nebula and the plerion around B1706-44 -- may serve as 
standard candles, with `grade A' according to \cite{weekes}. A third  plerion -- around the Vela pulsar -- was
given `grade B' in the same ranking.
All 10 gamma-ray pulsars are strong X-ray emitters. 

\begin{figure}
\vspace*{-0.7cm}
\begin{center}
\includegraphics[width=0.67\textwidth]{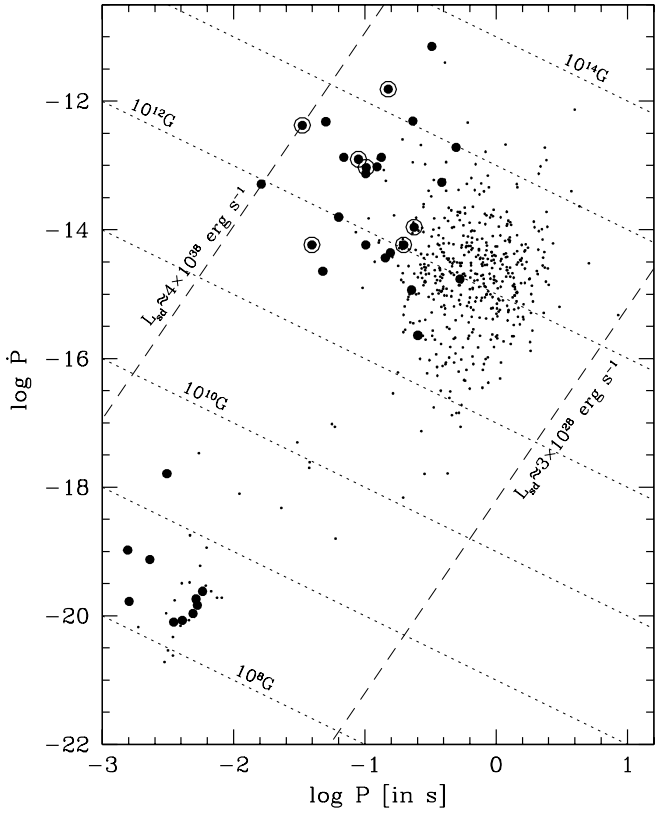}
\end{center} 
\vspace*{-2.5cm}
\caption[]{$P-\dot P$ diagram for Rotation Powered Pulsars. The pulsars detected exclusively in radio
are indicated with dots; they are taken mostly from the data base of \cite{taylor}. 
Thirty five pulsars emitting X-rays are indicated with bullets.
These include two objects recently discovered with RXTE: J0537-6910  
in SNR N157B in LMC  \citep{marshall} is the fastest young pulsar known, spinning 
twice as fast as the Crab pulsar but with
similar value of spin down luminosity; J1846-0258 
in SNR Kes-75 \citep{gotthelf}, with $P = 0.32$s, the highest $\dot P$ 
among all RPP. Seven bullets in circles indicate
seven gamma-ray pulsars of high confidence. Dashed lines correspond to constant values of spin down luminosity $L_{\rm sd}$.
The upper line ($L_{\rm sd}\approx 4\times 10^{38}$erg~s$^{-1}$) includes the Crab pulsar and J0537-6910,
the lower one ($L_{\rm sd}\approx 3\times 10^{28}$erg~s$^{-1}$) includes J2144-3933 -- the slowest ($P = 8.5$s)
radio pulsar detected so far \citep{young}. Dotted lines correspond to constant values of the 
dipolar component of the surface magnetic field as inferred from $P$ and $\dot P$ through (\ref{b6}) with $R_6 = 1$.
}
\label{ppdot}
\end{figure}

\begin{figure}
\vspace*{-0.7cm}
\begin{center}
\includegraphics[width=0.67\textwidth]{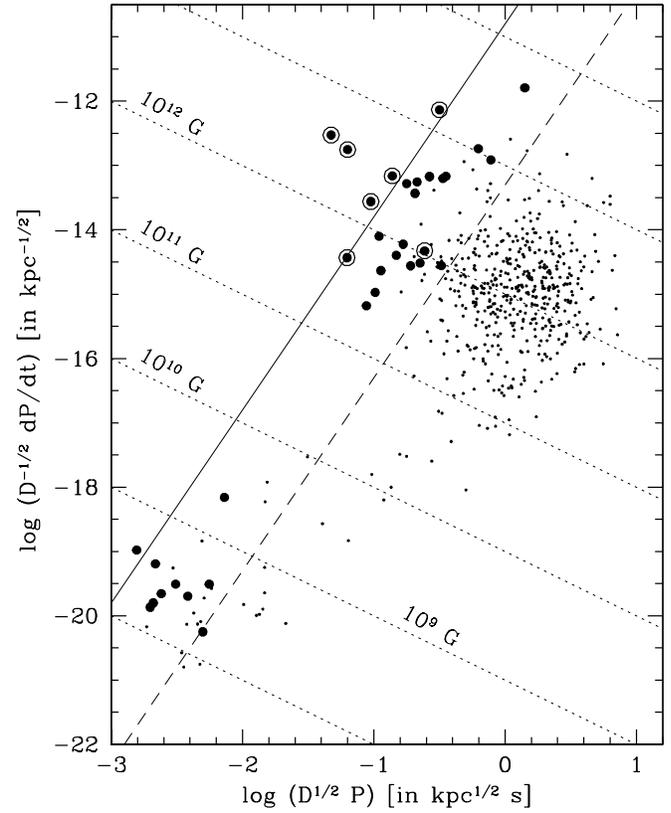}
\end{center} 
\vspace*{-2.5cm}
\caption[]{The ranking of RPPs by their spin-down flux $L_{\rm sd}/D^2$. The units of abscissa and 
ordinate are chosen to keep the resemblance with the  $P-\dot P$-diagram: the lines of constant spin-down
flux (two such lines - continuous and dashed -- are shown as examples) have the same slope with respect to the abscissa
($+3$) as the lines of constant \edot~ in Fig.~\ref{ppdot}; the lines of constant $\bs$ (dotted) 
have the same slope ($-1$) with respect to the abscissa in both figures; the range of 
values for coordinates is also identical.
Crab and Vela sticking out well above the continuous line
posses the highest spin-down fluxes among the RPPs. The continuous line is trained on B1951+32
with $L_{\rm sd}/D^2 \approx 7\times 10^{-8}\ergscm2$; the dashed line is for the millisecond X-ray pulsar
J0751+1807
with $L_{\rm sd}/D^2 \approx 2\times 10^{-10}\ergscm2$.
Note a low overall position of the gamma-ray pulsar B1055-52. 
(From \cite{kluzniak}, with recent updates).
}
\label{ranking}
\end{figure}

\begin{figure}
\vspace*{-0.7cm}
\centerline{\psfig{file=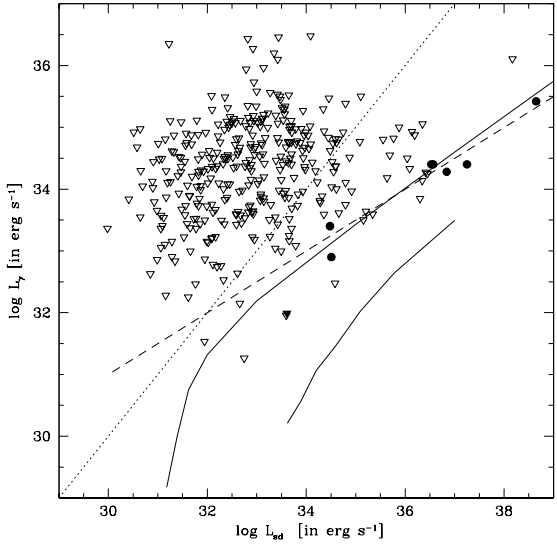,width=8.8cm} }
\vspace*{-0.7cm}
\caption[]{Gamma-ray luminosity versus
spin-down luminosity for seven pulsars (filled dots) detected with the CGRO instruments.
Opening angle of one steradian was assumed for the gamma-ray emission.
Open triangles are the EGRET upper limits after \cite{nel} for 350 objects, including
seven millisecond pulsars.
The filled triangle indicates the upper limit for J0437-4715 \citep{fierro95}. Note, that most of the upper
limits are well above the maximum possible value for $L_\gamma$ set by $L_\gamma = L_{\rm sd}$ (dotted line).
The dashed line marks the empirical relation derived for the CGRO pulsars:
$L_\gamma \propto L_{\rm sd}^{1/2}$. Solid
lines show evolutionary tracks for a classical pulsar with $B_{\rm s} = 10^{12}$G (upper line)
and a millisecond pulsar with $B_{\rm s} = 10^9$G (lower line)
according to the phenomenological model of \cite{rudak98}. 
}
\label{gamma}
\end{figure}

The positions of these HE pulsars are shown in the $P - \dot P$ diagram of Fig.\ref{ppdot} along with 
positions of about 700 radio pulsars for which $\dot P$ values were available. A remarkable fact is that the location
of X-ray sources does not correlate with the inferred  strength of magnetic field $B_{\rm s}$; at least not in 
a naively anticipated way that high-B objects would emit HE radiation, whereas low-B objects would not. In particular,
10 millisecond pulsars -- about thirty percent of all millisecond pulsars 
(the objects with $P \simless 0.01\s$ and $\dot P \simless 10^{-17}$, i.e. with low B values:
$B_{\rm s} \simless 10^9\G$) known to date -- have been detected as X-ray sources.
So far, millisecond pulsars have eluded the detection in gamma rays (possibly with one exception, see below) and
just upper limits were available for a handful of them from EGRET observations \citep{nel}.
In the case of J0437-4715 the upper limit is interestingly tight -- in disagreement
with the empirical relation $L_\gamma \propto L_{\rm sd}^{1/2}$ (see Fig.\ref{gamma}). 

Empirical evidence that pulsars must somehow tap their high-energy activity from rotation
comes from a strong correlation between 
the success of detection in X~and/or $\gamma$-rays and 
the position in the lists of targets ranked by 
spin-down flux values $L_{\rm sd}/D^2$, which is presented in a graphical form in Fig.~\ref{ranking}.
  
Spectral analysis for pulsars detected with ROSAT PSPC ($0.1\,\keV$ to $2.4\,\keV$) shows that
in most cases a power-law spectral model provides acceptable fits to the data \citep{becker97}.
Moreover, an intriguing empirical relation between inferred X-ray luminosity and spin down luminosity was found,
$L_X \approx 0.001\,L_{\rm sd}$, confirming rotational origin of most of the X-ray activity.
An interesting point is that the relation was obtained for all the sources regardless their 
temporal characteristics
(about 50~\% of all pulsars detected with ROSAT are unpulsed sources).
A complementary empirical relation was found for pulsed emission from 19 pulsars observed 
with ASCA ($0.6\keV$ to $10\keV$).
Assuming opening angle of X-rays to be one steradian,
the inferred pulsed X-ray luminosity correlates with spin-down luminosity as 
\begin{equation}\label{Lx}
L_X =  10^{34}\left(\frac{L_{\rm sd}}{10^{38}\,\ergs}\right)^{3/2}\ergs,  
\end{equation}
according to \cite{saito}.

A similar power-law empirical relation holds for gamma-rays (cf. Fig.\ref{gamma}),
but with a different power-law index (e.g. \cite{thompson1055}):
\begin{equation}\label{Lg}
L_\gamma \simeq 10^{35}\left(\frac{L_{\rm sd}}{10^{38}\,\ergs}\right)^{1/2}\ergs.
\end{equation}
  
Important conclusion from (\ref{Lx}) and (\ref{Lg}) is that  neither $L_X$ nor $L_\gamma$
becomes a sizable fraction of $L_{\rm sd}$. The most efficient conversion 
of spin-down luminosity into high-energy radiation is taking place for 
B1055-52 -- the oldest pulsar among {\it Seven Samurai} -- with $L_\gamma \simeq 0.1 \, L_{\rm sd}$. 

High energy lightcurves differ significantly from those in radio.  
Their most striking feature are relatively long duty cycles as well as phase shifts in comparison to the radio pulses.
Only for the Crab pulsar the peaks in gamma-rays as well as in radio wavelengths occur at the same
rotational phases.
The light-curve shapes fall into two categories. The Crab pulsar, Vela and Geminga
show two sharp peaks separated in phase by $0.4 - 0.5$ and connected
by an interpeak bridge of considerable level.
B1706-44 shows two peaks separated by 0.2 in phase, with some hints
of a third component in between. Other pulsars exhibit broad single pulses. 
Unknown opening angles for gamma-ray emission introduce a factor of uncertainty when inferring 
the gamma-ray luminosities. Broad peaks in gamma-ray pulses do not necessarily mean
large opening angles for gamma-ray emission. Polar cap
models, which rely on purely dipolar magnetic fields postulate nearly aligned rotators,
where inclination of magnetic axis to spin axis is  
comparable to the angular extent of the polar cap \citep{dau94}.

With a dicovery of sharply peaked pulsed X-ray emission in the fastest millisecond pulsar B1937+21
an apparently separate group of millisecond X-ray pulsars emerges,
with its members -- B1821-24 \citep{saito1}, J0218+4232 \citep{kuiper_milli}, 
and B1937+21 \citep{takahashi} -- being scaled-down versions of the Crab 
pulsar as far as sharp pulse profiles and
hard power-law X-ray spectra are concerned. An astonishing common feature within the group is the same strength
of  the magnetic field estimated at the light cylinder and the fact that it matches the strength of the Crab pulsar 
magnetic field at the light cylinder.

It is, however, lightcurves and broadband energy spectra 
extending from optical to X-rays and to gamma-rays, 
available for {\it Seven Samurai} (e.g. \cite{thompson1055}, \cite{thompson2001})
that are particularly impressive and challenging for any models of pulsar activity. 
These observations are reviewed
in details by \cite{kanbach2002} (this volume).

\section{Unipolar induction}
Consider a neutron star of radius $R_{\rm s}$ and surface magnetic field $B_{\rm s}$
as a perfect conductor.
For the rotating star with its dipolar magnetic field immersed in a vacuum 
an external quadrupole electric field $\vec {\cal E}$ develops, with non-zero
component along magnetic field lines at the surface.
The corresponding electrostatic potential $\Phi$  in polar coordinates $r$ and $\theta$ reads \citep{michel}
\begin{equation}\label{toy1}
\Phi (r,\theta) = - {Q \over r^3} \, \left(3 \cos^2\theta - 1 \right),
\end{equation}
where $Q = \pi\, \bs\,\rns^5/(3\, c\,  P)$ is the quadrupole moment.
Maximal electromotive force will then be induced between the pole of the star and the outer rim of the polar cap due to: 
\begin{equation}\label{toy4}
\Delta\Phi_{\rm pc} = \Delta\Phi_{\rm equator} \left(\frac{R_{\rm pc}}{R_{\rm s}}\right)^2,
\end{equation}
where 
\begin{equation}\label{toy2}
\Delta\Phi_{\rm equator} = \frac{1}{2c}\, B\, \Omega
\, R^2_{\rm s} 
\end{equation}
denotes a (huge but unavailabale) potential drop between the pole and the equator.
In terms of voltage, polar caps offer then
$\Delta V_{\rm pc} \approx 7 \times 10^{12}\, B_{12}\, P^{-2}\, {\rm Volt}$

\subsection{Potential drops in SCLF gaps}\label{drops}
Potential drop (\ref{toy4}) sets a firm upper limit for
all (model-dependent) realistic potential drops $\Delta\Phi_\|$, including  vacuum gap model of \cite{ruderman},
as well as outer gap models of e.g. \cite{ho}. 
In fact, 
\begin{equation}\label{drop}
\Delta\Phi_\| \ll \Delta\Phi_{\rm pc}
\end{equation}
for most cases.
We will concentrate hereafter on a class of models where the supply of
charged particles from a neutron star surface along open field lines 
is not limited by either binding or cohesive energy of the particles
and therefore can reach the Goldreich--Julian rate close to the surface   
(`Space Charge Limited Flow' --
\cite{sturrock}, \cite{arons79}). 
A concise but to-the-point account of essential properties of the SCLF models 
has been presented by \cite{arons00}. 

Three boundary conditions essential to the electrodynamics above the polar cap are 
(Muslimov \& Tsygan \cite{tsygan}):\\
1) $\vec {\cal E} \cdot \vec B = 0$ for the magnetosphere within the closed field lines,\\
2) $\Phi = 0$ at the surface and at the interface between the closed
magnetosphere and the open field lines,\\
3) ${\cal E}_\parallel = 0$ at the surface level,\\
where ${\cal E}_\parallel$ is the electric field component parallel to the magnetic field.\\
Last but not least, it is assumed that the outflow is stationary and the magnetosphere remains axisymmetric.

The electric field $\vec {\cal E}$ required to bring a charged particle into corotation satisfies the following equation
\begin{equation}\label{1}
  \vec {\cal E} + \frac{1}{c} ((\vec \Omega - \vec \omega_{\rm LT})
   \times \vec r) \times \vec B = 0
\end{equation}
where the inertial-frame dragging effect is included (\cite{beskin}, Muslimov \& Tsygan \cite{tsygan}) with  
$\vec \omega_{\rm LT} = \kappa_{\rm g} (R_{\rm s}/r)^3\,\vec \Omega$, where
$\kappa_{\rm g} = I/(M_{\rm s}R_{\rm s}^2)\cdot R_{\rm g}/R_{\rm s}$, and $R_{\rm g} = 2G M_{\rm s}/c^2$.

Charge density necessary to support this local $\vec {\cal E}$ is
\begin{equation}\label{2}
\varrho_{corot} = \frac{1}{4\pi}
\vec \nabla  \cdot \vec {\cal E} \approx -\frac{\vec \Omega \cdot \vec B}{2\pi c}\, 
\left[1 - \omega_{\rm LT} \right]
\end{equation}
The charge density $\varrho_{corot}$ due to SCLF at $r = R_{\rm s}$ is called the Goldreich--Julian charge 
density:
\begin{equation}\label{4}
\varrho_{\rm GJ} \approx -\frac{\vec \Omega \cdot \vec B}{2\pi c}\,\, [1 - \kappa_{\rm g}].
\end{equation}
As charged particles flow out along the open field lines a deviation of the local
charge density $\varrho_{\rm local}$  from the local corotation density $\varrho_{corot}$ develops.
By using now two relations satisfied 
in the dipolar structure, $B(r) \propto r^{-3}$ and $\varrho (r) \propto r^{-3}$, one obtains 
a simple formula for local deviation from the corotation charge density:
\begin{equation}\label{5}
\varrho_{\rm local} - \varrho_{corot} \approx \frac{\vec \Omega \cdot \vec B}{2\pi c}\,\, \kappa_{\rm g}\,\, \left[1 -
\left(\frac{R_{\rm s}}{r}\right)^3\right].
\end{equation}
Accordingly, the accelerating potential drop in SCLF reads
\begin{equation}\label{3}
\Delta\Phi_\| \approx \Delta\Phi_{\rm pc}\, \,\kappa_{\rm g} \, \left[1 -
\left(\frac{R_{\rm s}}{r}\right)^3\right],
\end{equation}
where $\Delta\Phi_{\rm pc}$ is given by (\ref{toy4}) and $\kappa_{\rm g} = 0.15\, I_{45}$.
This is a remarkable result obtained by Muslimov \& Tsygan~\cite{tsygan}: 
the potential drop due to inertial-frame dragging is significantly larger 
than the drop induced by curvature of magnetic field lines (`flaring', 
\cite{arons79}), or -- useless to say -  by particle inertia \citep{michel74}).

\section{Electric field structure in SCLF gaps}\label{efield}
The model with frame dragging effects  presented in a simplified form
above, does not take into account possible feed-back effect due to
$e^\pm$-pairs formed via photon absorption within open magnetic field lines. Copious pair creation occurs 
in a relatively thin
layer -- a pair formation front (PFF). The creation of pairs leads 
to the screening of
the accelerating field ${\cal E}_\|$ within the layer of PFF.
A detailed picture of this effect would require to follow the dynamics of electrons and positrons
in a self-consistent way. Instead, it is reasonable to assume, that  the field is shorted out
at the height were the first $e^\pm$-pair is created (hereafter denoted as $h_c$): ${\cal E}_\| = 0$ for $h \ge h_c$.

The problem of electric field structure in the context of SCLF with  boundary condition ${\cal E}_\| = 0$ set at
$h_0 = 0$ (stellar surface) and at $h \ge h_c$ (PFF)
was formulated and solved by \cite{hm98a}.
The solution is rather lengthy and includes special functions. It is however possible to
obtain simple but quite accurate analytical approximations. 
As long as the length $h_c$ of the accelerator is of the order of polar cap radius
$R_{\rm pc}$, the accelerating electric field 
may be approximated according to \cite{dr2000a} as
\begin{equation}
{\cal E}_\parallel \approx -1.46\ \frac{B_{12}}{P^{3/2}}\ h \left(1 - \frac{h}{h_c}\right) 
f_1(\xi) \cos\chi\ {\rm Gauss},
\label{electric1}
\end{equation}
where
$B_{12}=B_{\rm s}/10^{12}\G$, $P$ is the spin period in seconds, $\chi$ is the angle between
the spin axis and the magnetic moment of the rotating star, $h$
is expressed in cm, and $M=1.4\, M_\odot$, $R_{\rm s}=10^6$cm.
The magnetic colatitude $\xi\equiv \theta/\theta(\eta)$ 
is scaled with the half-opening angle of the polar magnetic flux tube $\theta(\eta)$, 
where $\eta\equiv 1 + h/R_{\rm s}$. The magnetic colatitude function $f_1(\xi)$ is a monotonically
decreasing function, with $f_1(0) \simeq 1$ and  $f_1(1) = 0$.

Vertical structure of the electric field depends (via the location of PFF) on radiative processes
which induce the pair creation: curvature radiation (CR) and
inverse Compton scattering (ICS) on soft X-ray photons from the stellar surface (brief characteristics 
of these processes is presented in the next section).

An interesting way to avoid immediate screening of the electric field by created pairs,
relevant in strong magnetic fields $\ga 0.1\, \bcr$, 
was proposed by \cite{usov95} who noticed that pairs are created then near
the energy threshold,  most favourably
in bound states -- as atoms of positronium -- which then move to high altitudes before being ionized.


\section{Radiative processes in pulsar magnetospheres}\label{radiative}
Cooling of ultrarelativistic electrons via curvature radiation (CR) and magnetic inverse Compton scattering (ICS) 
are the most natural ways of producing hard gamma-rays capable of inducing
cascades of $e^\pm$-pairs and secondary HE photons.
These two processes dominate within two distinct ranges of Lorentz factors $\gamma$ of primary electrons.

When $\gamma \simless 10^6$, magnetic inverse Compton scattering  plays a dominant role in braking the electrons 
and it is the main source of hard gamma-ray photons \citep{sdm}. 
Energy losses due to resonant ICS limit the Lorentz factors of the particles to a level 
which depends on electric field strength $\cal E_\parallel$, temperature $T$ and size of 
hot polar cap, and magnetic field strength $B_{\rm s}$ (\cite{xia}, \cite{sturner}).
The Lorentz factors can then be limited even to $\sim 10^3$. 
This stopping effect becomes more efficient for stronger
magnetic fields, and it was suggested as an explanation for 
the observed cutoff at $\sim 10\MeV$ in the spectrum of B1509-58 \citep{sturner}. ICS was also found to
be able (under some circumstances) to smear out the monotonic energy distribution of electrons significantly, with
interesting consequences for the cascades \citep{dr2000b}.

However, in their modern versions the accelerators of particles are strong enough to outpower the ICS cooling.
In consequence, very high Lorentz factors -- $\gamma\simgreat 10^6$ -- are achieved by electrons, limited by CR.  
The first detailed scenario of radiative processes in CR-induced cascades was presented by \citep{dau82} and
despite many modifications and additions its basic features remain valid. The model assumes
that primary electrons accelerated to ultrarelativistic energies emit curvature photons which in turn are
absorbed by the magnetic field and $e^\pm$-pairs are created. These pairs cool off instantly
via synchrotron radiation process. Whenever the SR photons are energetic enough they may
lead to further creation of pairs, etc.
ICS can still be incorporated to the models with CR-induced cascades as the process relevant for  
$e^\pm$-pairs, since typical Lorentz factors of theirs do not exceed $\sim 10^3$. 
According to the analytical model of \cite{zh2000a} the empirical relations
for X-ray and gamma-ray luminosities of pulsars (presented in Sect.3) can be reproduced satisfactorily 
when the ICS involving $e^\pm$-pairs is included.

Processes relevant for generation and transfer of high-energy photons in pulsar magnetospheres are therefore: \\
 - curvature radiation,\\
 - magnetic inverse Compton scattering, \\
 - magnetic pair creation \quad ($\gamma \, \rightarrow e^\pm$) \\
 - synchrotron radiation,\\
 - photon splitting \quad ($\gamma \rightarrow \gamma + \, \gamma$,)\\
 - photon-photon pair creation \quad ($\gamma + \, \gamma \rightarrow e^\pm$).\\
Their
basic properties are given below, except for $\gamma + \, \gamma \rightarrow e^\pm$:
in practice it is treated exactly as in free space. 
This is justified in models of `thick outer gaps' \citep{cheng} but within the framework
of polar cap models this is not the case, in general. 
However, no handy formula is available for the cross section of this process in the limit of high $B$ 
and standard non-magnetic formulae are in use (e.g. \cite{bing2}).

To illustrate how these processes contribute to the high-energy radiation of a pulsar
the numerically calculated components (due to the first four processes in the list) 
are presented in Figs.\ref{vela} and \ref{vela2} (after \cite{dr2002a} and \cite{drb2001}) 
along with overlaid data points for
the Vela pulsar. The dipolar field in the Vela pulsar does not exceed $10^{13}\G$ 
and photon splitting was neglected as non-competitive to the magnetic pair creation. 
The electric field structure of the accelerator used in these calculations is taken after \cite{hm98a}.

\subsection{Curvature radiation}
Relativistic electron of energy $\gamma m c^2$ (we take $\gamma \gg 1$) 
sliding along the magnetic field line
of curvature $\varrho_{\rm cr}$ will emit photons with a continuum  energy spectrum 
peaked at the characteristic energy
\begin{equation}
\varepsilon_{\rm cr} = {3 \over 2} \, c \, \hbar \, {\gamma^3 \over \varrho_{\rm cr}}.
\label{cr1}
\end{equation}
The radius $\varrho_{\rm cr}$
for a purely dipolar line attached to the outer rim of the polar cap can be approximated
not far away from the NS surface as  $\varrho_{\rm cr} \approx \sqrt{ R_{\rm s} \cdot R_{\rm lc}} \approx 10^8 \sqrt{P} \cm$.
The cooling rate of that electron is
\begin{equation}
\dot \gamma_{\rm cr} = - {2 \over 3} \, {e^2 \over m c}\, {\gamma^4 \over \varrho_{\rm cr}^2}.
\label{cr10}
\end{equation}
For a monoenergetic injection function of electrons $Q(\gamma) \propto \delta (\gamma - \gamma_0)$
and their cooling due solely to CR the electrons will assume a single power-law distribution
in energy space $N_{\gamma}({\rm el.}) \propto \gamma^{-4}$ for $\gamma < \gamma_0$ as long as they stay
within the region of the cooling. Their escape introduces a natural low-energy 
cutoff $\gamma_{\rm cutoff}$ in $N_{\gamma}({\rm el.})$.
Therefore, the unabsorbed CR energy spectrum $f_\varepsilon (\varepsilon)$ due to the injected 
electrons has a broken power-law shape, with a high-energy limit set by $\gamma_0$
and the break at some energy $\varepsilon_{\rm break}$.
For $\varepsilon > \varepsilon_{\rm break}$ the energy spectrum is
$f_\varepsilon (\varepsilon) \propto \varepsilon^{-2/3}$, and 
$f_\varepsilon (\varepsilon) \propto \varepsilon^{+1/3}$ for $\varepsilon < \varepsilon_{\rm break}$.
Since nonthermal spectra cover ususally many decades in energy it is more
convenient to use $\varepsilon f_\varepsilon (\varepsilon)$ 
for easy comparison of power in different parts of energy space
(see Fig. \ref{vela}).
Accordingly,
$\varepsilon f_\varepsilon (\varepsilon) \propto \varepsilon^{+1/3}$ above the break, and  $\propto \varepsilon^{+4/3}$
below the break.

The cutoff limit $\gamma_{\rm cutoff}$ can be found by comparing the characteristic cooling time scale
$t_{\rm cr} \equiv \gamma/ |\dot \gamma_{\rm cr}|$ with the estimated time of escape  $t_{\rm esc}$,
which we take as  $t_{\rm esc} \approx \varrho_{\rm cr}/ c$. 
Therefore 
\begin{equation}
\varepsilon_{\rm break} \approx  {9 \over 4} \, \hbar \, {c \over r_0} \approx 150\, \MeV,
\label{cr3}
\end{equation}
where $r_0$ is the classical electron radius (Rudak \& Dyks~\cite{rudak99}). Note, that 
photon energy $\varepsilon_{\rm break}$ at which the cooling break occurs does not depend on any
pulsar parameters. 

The spectrum of CR calculated  numerically to model the Vela pulsar \citep{drb2001}
is shown in Fig.\ref{vela} as dot-dashed line. 
High-energy cutoff due to one-photon magnetic absorption occurs around $10\,$GeV. Note the importance of 
gamma-ray detectors capable to operate above $10\,$GeV for (in)validating the model.
The low-energy CR spectral break $\varepsilon_{\rm break}$ is prominent at $\sim 40 \MeV$.
Below $\varepsilon_{\rm break}$ the power of CR decreases and eventually  
becomes unimportant at $\sim 1 \MeV$ where the synchrotron component takes over.

\subsection{Magnetic pair creation}
Pair creation via magnetic photon absorption ($\gamma + \vec B
\rightarrow e^\pm + \vec B$) is kinematically correct because the magnetic field
can absorb momentum. To ensure high chances for the process to occur it is not
enough for a photon propagating at a pitch angle $\psi$ to local $\vec B$
to satisfy 
the energy threshold condition, $\sin \psi \cdot \varepsilon \ge 2 m c^2$,
but high optical thickness $\tau_{\gamma B}$ of the magnetosphere is required.
In fact the condition $\tau_{\gamma B} = 1$ is used as a criterium for the so called death-line 
for radiopulsars in the $P - \dot P$ diagram. Maximal values of $\sin \psi$ for 
curvature photons in the dipolar field do not exceed
$\sim 0.1 \sin \theta_{\rm pc} \approx 0.0014\, P^{-1/2}$ \citep{sturrock},
and, therefore, both conditions are difficult to be met for long-period pulsars.

The absorption coefficient for the process as
described in \cite{erber} and used to calculate $\tau_{\gamma B}$  reads 
\begin{equation}
\eta(\varepsilon) = {1\over 2} {\alpha \over \lambdabar_{\rm c}} {B_\perp
\over B_{\rm crit}} T\left(\chi \right)
\label{mpp1}
\end {equation}
where $\alpha$ is a fine structure constant, $\lambdabar_{\rm c}$
is a Compton wavelength, 
$B_{\rm crit} = m^2 c^3/e\hbar \simeq 4.4\times 10^{13}\G$,
$B_\perp$ is the component of the
magnetic field perpendicular to the photon momentum, and $\chi
\equiv {1\over 2}{B_\perp \over B_{\rm crit} }{\varepsilon\over
m_e c^2}$ is the Erber parameter $\chi$.  The function $T(\chi)$ is then approximated as
 $T(\chi) \approx 0.46 \exp\left(-4 f/3\chi
\right)$, 
which is valid for $\chi \simless 0.2$; for $\chi \simgreat 0.2$
this approximation starts to overestimate $\eta(\varepsilon)$ significantly.
The function $f$ is a near-threshold correction introduced
by \cite{dau83}, important in the case of high local $B$. 
Electron-positron pairs created through the magnetic absorption
emit synchrotron photons (SR) and also may get involved in ICS with soft photons.

Taking advatage of the dipolar character of the magnetic field, it straightforward to find
analytical approximation of the condition
$\tau_{\gamma B} = 1$.
A photon created at a radial distance  $r$ from the neutron star center, with its momentum
parallel to the local magnetic field line 
will undergo magnetic absorption if its
energy exceeds the following local cutoff energy (after \cite{bh2000}, \cite{b2001})
\begin{eqnarray}
\varepsilon_{\rm cutoff} & \approx & 0.4 \, P^{1/2} 
\left( {r \over \rns} \right) ^{1/2} 
\nonumber\\
&& \times \, \max \left\{ 1, \, {0.1 \bcr \over \bpc } \left( {r \over \rns }\right) ^3\right\}  \GeV.
\label{cutoff1}
\end{eqnarray}
This formula is valid for the magnetic field lines with their footpoints at the outer rim of the polar cap,
of colatitude $\theta_{\rm pc}$ (eq.~\ref{pc}),
hence the dependence on $\sqrt P$. For other colatitudes $\theta$ the formula 
should be multiplied by $\theta_{\rm pc}/ \theta$.

\subsection{Synchrotron radiation}
Consider a particle of energy $\gamma m c^2$ gyrating around a local field line at a pitch angle $\psi$.
Let $\gamma_\parallel$ denotes the Lorentz factor of the reference frame comoving with the center
of the gyration. As long as $\gamma_\parallel \gg 1$  it relates to the pitch angle $\psi$
via $\sin \psi \approx \gamma_\parallel^{-1}$.
The energy available for synchrotron emission at the expense of the particle is  $\gamma_\perp m c^2$,
and $\gamma = \gamma_\perp \, \gamma_\parallel$. 

The rate of SR cooling reads
\begin{equation}
\dot \gamma_{\rm sr} = - {2 \over 3} \, 
{r_0^2 \over m_{\rm e} c}\, B^2 \gamma_\perp^2 = - {2 \over 3} \,
{r_0^2 \over m_{\rm e} c}\, B^2 \sin ^2\psi \,\gamma^2.
\label{sr0}
\end{equation}
In comparison to the CR cooling it is  enormous (due to much smaller curvature radius). 

Critical photon energy (analogous to (\ref{cr1})) reads 
\begin{equation}
\varepsilon_{\rm sr} = {3 \over 2} \, \hbar \, {e B \over m_{\rm e} c} \gamma^2 \sin \psi.
\label{sr1}
\end{equation}

For a monoenergetic injection function of particles ($e^\pm$-pairs in the context of this review) 
and their cooling due to SR the energy spectrum of SR  
spreads between a high-energy limit $\varepsilon_{\rm sr}(\gamma_0)$ set by 
$\gamma_0$ of the injected (created) particles, and 
a low-energy turnover $\varepsilon_{\rm ct}$ determined by the condition $\gamma_\perp \sim 1$:
\begin{equation}
\varepsilon_{\rm ct} \equiv \varepsilon_{\rm sr}(\gamma = \gamma_\parallel) 
= {3 \over 2} \, \hbar \, {e B \over m_{\rm e} c} \,{1 \over {\sin \psi}}.
\label{sr2}
\end{equation}
The spectrum assumes a single power-law shape $f_\varepsilon (\varepsilon) \propto \varepsilon^{-1/2}$
(and accordingly -- 
$\varepsilon f_\varepsilon (\varepsilon) \propto \varepsilon^{+1/2}$) above the turnover.
Below $\varepsilon_{\rm ct}$, the spectrum $f_\varepsilon$
changes it slope, asymptotically reaching $\propto \varepsilon^{+2}$.
It is built up by contributions from low-energy tails emitted by particles with $\gamma_\perp \gg 1$, 
and each low-energy tail cuts off at local gyrofrequency, which in the reference frame
comoving with the center of gyration is $\omega_B = {e B \over m_{\rm e} c \,\gamma_\perp}$.
The spectrum of SR calculated with Monte-Carlo method to model the Vela pulsar is shown in
Fig.\ref{vela} as a dashed line.  
The low-energy part of the SR spectrum at $\varepsilon_{\rm ct}$ seems to be essential for 
connecting the RXTE and the OSSE points.

Synchrotron as well as cyclotron emission become prohibited at strong local magnetic fields
$B \ga 0.1\, \bcr$ (\cite{usov95}, \cite{arendt}) -- the magnetic photon absorption proceeds here close to the
threshold, and the pairs created do not occupy excited Landau levels. In consequence,
electromagnetic cascades are expected to be notably weaker than for  $B < 0.1\, \bcr$.
Along with photon splitting \citep{bh2001}, this may explain why do radiopulsars avoid
the high-B region in the $P-\dot P$ diagram.

\subsection{Magnetic Inverse Compton Scattering}
Consider an electron with a Lorentz factor $\gamma$ moving along a magnetic field line $\vec B$
and a photon of energy $\varepsilon = \epsilon m c^2$ moving at angle $\arccos{\mu}$ to the
field line. 
In the reference frame comoving with the electron (primed symbols)
the counterpart of the free-space Compton formula, due to energy-momentum conservation
appropriate for collisions with the electron at the ground Landau
level both in the initial and final state, reads 
\begin{eqnarray}
\lefteqn{ \epr_s = \left(1 -
{\mu^\prime_s}^2\right)^{-1}\left\{\vbox{\vskip5mm} 
1 + \epr(1 - \mu^\prime\mu^\prime_s)\ + \right.}\nonumber\\
&&\hbox{\hskip4mm}\left.-\left[  1 + 2\epr\mu^\prime_s (\mu^\prime_s -
\mu^\prime) +
{\epr}^2 (\mu^\prime_s - \mu^\prime)^2  \right]^{1/2}\right\}
\end{eqnarray}
where $\epr=\epsilon\gamma(1 - \beta\mu)$ \citep{herold}, and
symbols with no subscript and  with the subscript `s' refer to the state before the scattering and 
after the scattering, respectively.
A longitudinal momentum of the electron 
in the electron rest frame changes due to recoil from zero to $(\epr\mu^\prime -
 \epr_s\mu^\prime_s)mc$.

The polarization-averaged relativistic magnetic
cross section in the Thomson regime may be approximated 
with a nonrelativistic formula \citep{dermer}:
\begin{equation}
\sigma = \frac{\st}{2}\left(1 - {\mu^\prime}^2 + (1 +{\mu^\prime}^2)
\left[g_1 + \frac{g_2 - g_1}{2}\right]\right)
\label{crosssection}
\end{equation}  
where $\st$ is the Thomson cross section, and $g_1$ and $g_2$ are given by
\begin{equation}
g_1(u) = \frac{u^2}{(u+1)^2}, \hbox{\hskip 1cm}
g_2(u) = \frac{u^2}{(u-1)^2 + a^2}
\end{equation}
where $u\equiv\epr/\eb$, $a\equiv 2 \alpha\eb/3$, $\eb \equiv \hbar e B/m^2c^3$ and $\alpha$ is a
fine-structure constant. 
The resonance condition for the scattering is therefore the cyclotron resonance $\epr = \eb$.
The factor $a$ represents a `natural' broadening of the resonance due to finite lifetime at the excited 
Landau level.

In the Klein-Nishina regime ($\epr > 1$) the relativistic magnetic cross 
section for the $|\mu^\prime|\approx 1$ case becomes better approximated with the well
known Klein-Nishina relativistic nonmagnetic total cross section $\skn$
(\cite{dau86}, \cite{dermer}).

The rate $\cal R$ of scatterings subject by an electron moving across the field of soft photons,
measured in the lab frame is
\begin{equation}
{\cal R} = c \int d\Omega \int d\eps\ \sigma
\left(\frac{dn_{\rm ph}}{d\eps d\Omega}\right)(1 - \beta\mu)
\label{rate}
\end{equation}
where $\Omega=d\mu d\phi$ is the total solid angle
subtended by the source of soft photons, $\mu = \cos{\theta}$, $\sigma$ is
a total cross section for the process, and $dn_{\rm ph}/d\eps/d\Omega$ 
is the local density of the soft photons.

The properties of the field of soft photons are usually simplified by
taking $dn_{\rm ph}/d\eps/d\Omega$ as for the blackbody radiation.
This simplification should be taken with care since magnetised atmospheres
of neutron stars introduce
strong anisotropy as well as spectral distortions to the outgoing radiation \citep{pavlov}.
Effectively it means that the ICS effects obtained with this simplification are just upper limits
to the actual effects.

To estimate electron cooling rate $\dot{\gamma}_{\sss \rm ICS}$  due to the ICS the differential 
form of (\ref{crosssection}) is necessary:
\begin{eqnarray}
\lefteqn{\frac{d\sigma}{d\Omega^\prime_s} = \frac{3\st}{16\pi}
\,
\left[
(1 - {\mu^\prime}^2)(1 - {\mu^\prime_s}^2)\ + \right.}\nonumber\\
&&\hbox{\hskip12mm}\left.+\ \frac{1}{4}\ (1 + {\mu^\prime}^2)
(1 + {\mu^\prime_s}^2)(g_1 + g_2)\right]
\label{diffsection}
\end{eqnarray}
(eg. \cite{dau89}), where $d\Omega^\prime_s = d\phi^\prime_s d\mu^\prime_s$ is 
an increment of solid angle into which outgoing
photons with energy $\epr_s$ in the electron rest frame are directed. 
The mean electron energy loss rate then reads
\begin{eqnarray}
\lefteqn{\dot{\gamma}_{\sss \rm ICS} = - c\int
d\epsilon \int d\Omega\, \left(\frac{dn_{\rm ph}}{d\epsilon d\Omega}\right)
(1 - \beta\mu)\ \times}\nonumber\\
&&\hbox{\hskip20mm}\times 
  \int
d\Omega^\prime_s
\left(  \frac{d\sigma}{d\Omega^\prime_s}  \right)
(\epsilon_s - \epsilon)
\label{lossrate}
\end{eqnarray}
where $\epsilon_s = \epr_s\gamma(1 + \beta\mu^\prime_s)$ is the scattered
photon energy in the lab frame (e.g. \cite{dermer}). 

\begin{figure*}
\centerline{\psfig{file=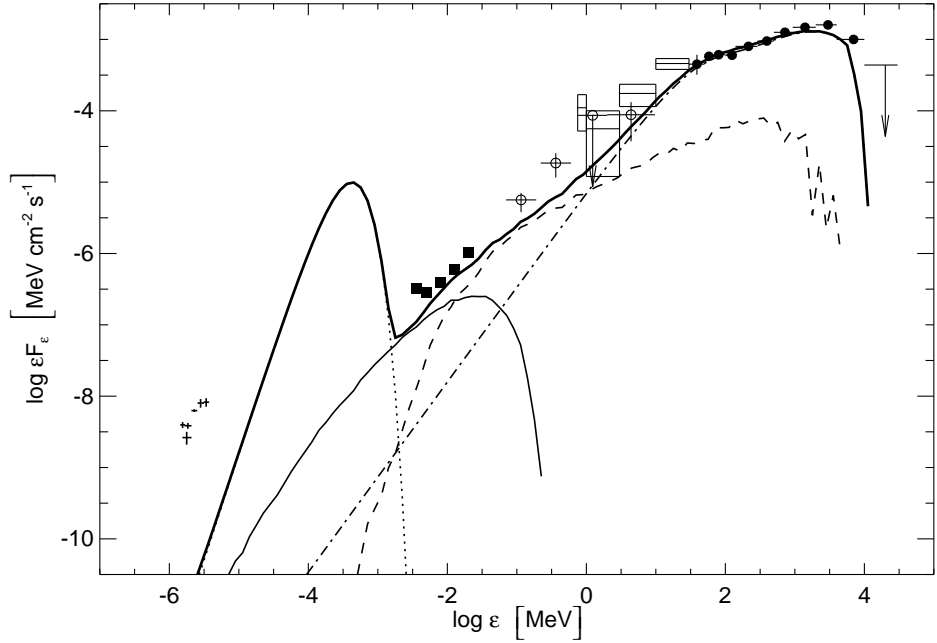,width=13.cm} }
\caption[]{The model energy spectrum calculated by \cite{dr2002a} to reproduce the spectral 
features of the Vela pulsar ($P=89$ ms, $B_{\rm s} = 6\times 10^{12}\G$).
The accelerator is located at $h_0 = 3 \rns$ above the surface (see Sect.6).
The broad--band spectrum consists of four components due to:
curvature radiation of primary electrons (dot-dashed), synchrotron radiation of
secondary $e^\pm$--pairs (dashed), inverse Compton scattering of surface X-ray photons on
the $e^\pm$--pairs (thin solid) and the blackbody surface emission (dotted). 
The surface temperature $T_{\rm s} = 1.26\times 10^6$K was assumed for the neutron star.
Total spectrum  is given by a thick solid line.
Phase-averaged data points for Vela from different telescopes are indicated:
crosses -- multicolor photometry with NTT and HST \citep{vela_optical};
filled squares -- RXTE \citep{vela_rxte}; open circles -- OSSE \citep{vela_osse};
rectangles -- COMPTEL \citep{comptel_1st_cat}; filled circles plus an upper 
limit just above $10$~GeV -- EGRET \citep{cgro}. 
}
\label{vela}
\end{figure*}

\begin{figure*}
\centerline{\psfig{file=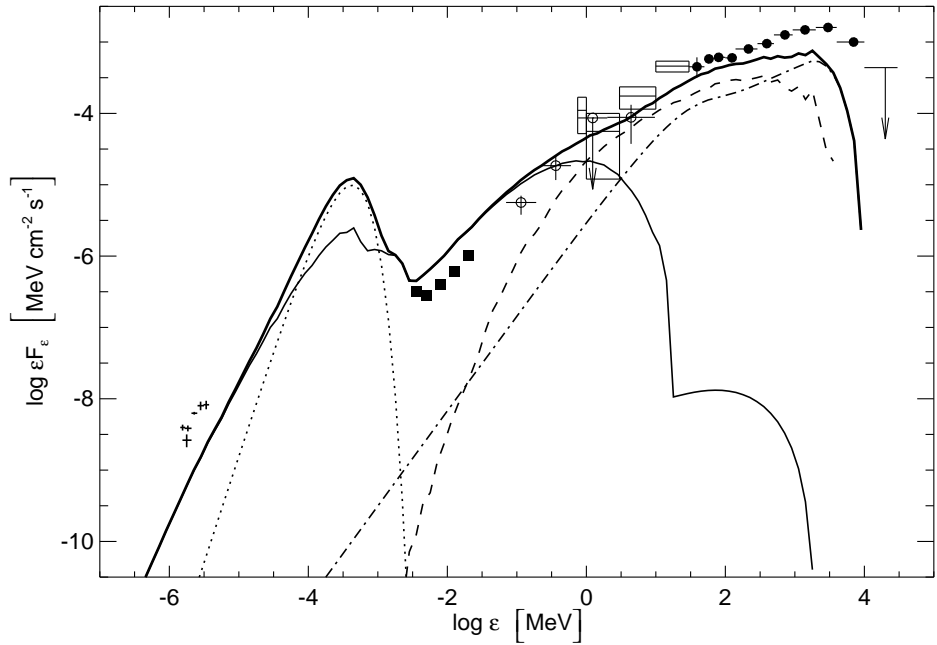,width=13.cm} }
\caption[]{Same as Fig. \ref{vela} but with two parameters changed
drastically ($B_{\rm s} = 0.6\times 10^{12}\G, \ h_0 = 0$) in order to 
obtain as much power in the optical range as possible by
modifying the ICS component due to $e^\pm$--pairs (thin solid line).
Despite
unrealistically favourable conditions to down-scatter some of the blackbody photons,
the low-energy tail of the ICS component falls short by a factor of $\sim 3$ below
the observed optical flux level (crosses).
}
\label{vela2}
\end{figure*}

The spectrum of magnetic ICS calculated  numerically to model the Vela pulsar 
is shown in Figs.\ref{vela} and \ref{vela2} with thin solid line. The blackbody soft photons originating
at the stellar surface (dotted line) are upscattered by secondary $e^\pm$-pairs at the expense of their
``longitudinal" energy $\gamma_\parallel m c^2$, assumed to remain unchanged during
the burst of synchrotron emission.
Without the ICS component due to the $e^\pm$-pairs the RXTE data for the Vela pulsar would be difficult
to reproduce within the model.
It is worth to note, that the magnetic ICS component due to primary electrons (not shown in 
Fig.\ref{vela}) is energetically insignificant comparing
to the CR component.
  
\subsection{Photon splitting}
Photon splitting into two photons in the presence of magnetic field $B$  is a third-order
QED process with no energy threshold \citep{adler}. 
The attenuation coefficient, after averaging over the polarization states, reads \citep{splitt97}
\begin{equation}
T_{\rm split}(\epsilon) 
\approx \frac{\alpha^3}{10\pi^2} \frac{1}{\lambdabar_{\rm c}} \left(\frac{19}{315}\right)^2 
\left(\frac{B \sin{\theta_{\rm kB}}}{B_{\rm crit}}\right)^6 \,  \epsilon^5 \, \,  {\rm cm}^{-1},
\label{split1}
\end{equation}
(provided $B$ does not exceed $B_{\rm crit}$ substantially)
where $\alpha$ is a fine structure constant, $\epsilon$ is the photon energy in units of $m c^2$,
$\theta_{\rm kB}$ is the angle between photon momentum vector and the local magnetic field.  
The process, therefore, strongly depends on magnetic field strength $B$.

Photon splitting has attracted substantial interest in recent years due to the discovery of
neutron stars with supercritical magnetic fields \citep{mereghetti}.
It has been analysed in details by \cite{splitt97} and incorporated in a Monte Carlo code
tracing the propagation of electromagnetic cascades in the magnetospheres of high-$B$ pulsars. 
The effect was found to explain satisfactorily the unusual cut-off observed in the gamma-ray spectrum of
B1509-58 (see Sect.3). Generally, it becomes competitive to the magnetic pair creation
for dipolar magnetospheres with $B_{\rm s} \simgreat 0.3 B_{\rm crit}$ \citep{splitt97}. The degradation of photon energy
in the course of splitting inhibits also any development electromagnetic cascades.
In consequence, high-$B$ RPP should not emit coherent
radio emission. 
Indeed, there exists a high-$B$ region in the $P-\dot P$ diagram void
of radiopulsars. Even though three recently discovered 
(during The Parkes Multibeam Pulsar survey) high-$B$ radiopulsars 
\citep{camilo} are located above
the formal limiting line derived in \cite{bh98} (and elaborated in \cite{bh2001}), 
the general argument that magnetars are expected to be
radio-quiet RPP remains valid \citep{zh2000b}.  


\section{Facing the Vela pulsar}\label{problems}
In order
to fit our model
to the phase averaged spectrum of Vela over almost 8 decades in energy (see Fig.\ref{vela}) 
we were forced to locate the accelerator at the altitude of $h_0 \simeq 3\, \rns$, i.e. unrealistically
high for any version of polar-cap models. The same problem was encountered by \cite{dau96} when modelling
the CGRO data points of Vela.

Physical justification for lifting the accelerator up to $\sim 1\,\rns$ might rely on a mechanism proposed by
\cite{hm98a}. They considered a sandwich-like structure of two pair formation fronts (PFF),
controlled by either CR or ICS, where the lower PFF would be formed due to  positrons returning 
from the upper PFF.
With reasonable surface temperatures ($T_{\rm s} \sim 5\times 10^5\K$) 
it is ICS which dominates (comparing to CR) in generating 
high-energy quanta at the energetic expense of
upward moving electrons and downward moving positrons.
Such a situation is not symmetrical and not stable, therefore. Accordingly, the lower PFF tends to elevate until
the CR-cooling takes over, which happens at $h_0 \sim 1\, \rns$.
Unfortunately, this mechanism -- rather promising in the context of
required $h_0 \simeq 3\, \rns$, even though $3\, \rns$ is still notably more than $\sim 1\, \rns$ -- has been
invalidated by recent numerical calculations:  \cite{hm2002} 
concluded  the inability of ICS-controlled upper PFFs  to develope 
lower PFFs because of too low multiplicities in such cases.

\begin{figure}
\vspace*{-1.3cm}
\begin{center}
\includegraphics[width=0.63\textwidth]{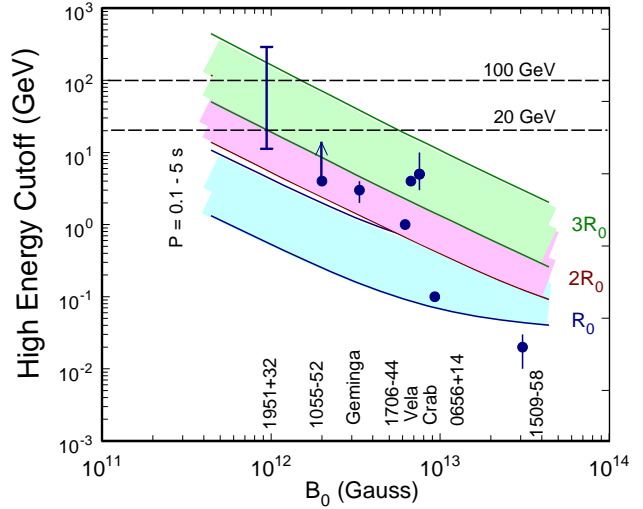}
\end{center}
\vspace*{-1.cm}
\caption[]{Spectral cutoff energies estimated for eight gamma-ray pulsars are plotted against inferred surface
magnetic field strengths $\bs$. Three shaded bands (note the overlaps) indicate 
the location of $\varepsilon_{\rm cutoff}$
according to eq.(\ref{cutoff1}) for three values of the emission altitude ($r = \rns, \, 2\rns$ and $3\rns$)
and for the range of spin periods $P$ between $0.1$ and $5\, \s$.
(From \cite{bh2000} and \cite{b2001}; courtesy of Matthew Baring.)}
\label{baring}
\end{figure}

The requirement of high-altitude accelerator for Vela is likely to be extended to other
gamma-ray pulsars. 
A piece of hint for such a requirement comes from   
a strong correlation for known gamma-ray pulsars between the inferred magnetic field 
strength and the value of high-energy cutoff \citep{bh2000}.
[\emph {Note, that this empirical  correlation speaks in favor of 
high-energy cutoffs in pulsar spectra
originating due to magnetic absorption 
of gamma-rays -- a signature of polar-cap models}.]
This is shown 
in Fig.\ref{baring} along with the dependence of $\varepsilon_{\rm cutoff}$ on the
location of photon emission point $r$ 
according to eq.(\ref{cutoff1}).

Last but not least -- a real  challenge for all contemporary 
models for Vela comes from the phase-resolved spectral analysis
of pulsed hard X-rays 
and the phasing of the lightcurves in optical, X-rays and gamma-rays, 
presented by \cite{rosetta1} and \cite{rosetta}.
The strong double-peak lightcurve in optical (fig.1 in these refs.) is particularly enigmatic.
The peak-to-peak separation is notably smaller than
in HE domain. Moreover, the leading peak has no 
counterpart in any other
energy range (including the radio). The trailing optical peak seems to be associated 
with the trailing RXTE peak
and with the trailing gamma-ray peak, though the latter lags behind the trailing optical by about $0.1$ in phase. 
Extrapolation of the gamma-ray spectrum taken for 
either of the two gamma-ray peaks falls below the phase-averaged
optical flux level (by a factor $\sim 10$ in the case of the trailing gamma-ray peak).

The problem of optical emission is also difficult to solve 
in the framework of our phase-averaged approach to model the spectrum of Vela.
A difficult task we have faced was to find
a  component (without referring to any ingredients external to the model)
that would account for
the optical emission, at least in some particular circumstances. 
Fig.\ref{vela2} gives an example of what can be achieved at best in the optical domain
of the model spectrum, but at the price  paid at the high-energy end of the spectrum.

\section{Millisecond pulsars}\label{milli}  
\begin{figure}
\begin{center}
\includegraphics[width=0.4\textwidth]{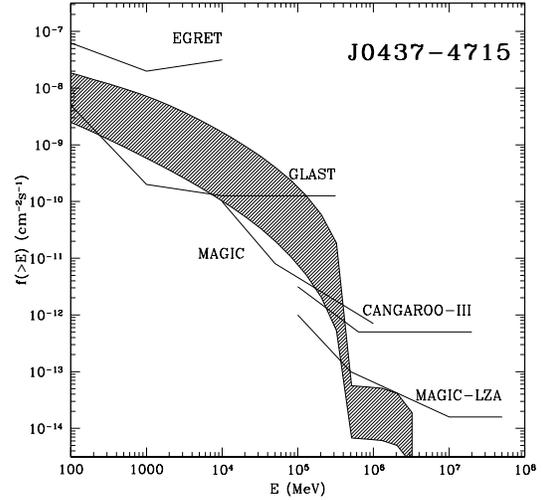}
\end{center} 
\vspace*{-0.6cm}
\caption[]{Cummulative~spectral flux of photons expected for the millisecond pulsar
J0437-4715 at a distance of $140\,$pc. 
The shaded region shows the range of flux levels due to uncertainity
in the maximal energy of primary electrons. The main part of the spectrum is due
to curvature radiation of the electrons. The additional feature reaching the VHE
domain is due to inverse Compton scattering of soft photons 
from the surface with the temperature $4\times 10^5$K.
Sensitivities of EGRET as well as three major HE and VHE 
experiments of the future are also indicated. MAGIC--LZA denotes
sensitivity of MAGIC in its Large Zenith Angle mode. (After \cite{bulik}). 
}
\label{j0437}
\end{figure}

The question of whether millisecond pulsars (i.e. with low magnetic fields)
do emit gamma-rays will be hopefuly answered within the next five years with
GLAST.
On theoretical
side, there had been some attempts to address this problem. For instance, \cite{wei}  presented
broad-band spectra of both pulsed and unpulsed emission expected
in their version of the outer-gap model.  
In the
framework of polar-cap scenarios, 
\cite{sd94} made their predictions for gamma-ray
luminosity \Lg by scaling down with $B$ their analytical formulae for \Lg for classical pulsars.  
\cite{rudak98} proposed a modification of the
model by \cite{dau82}; this modification incorporates a contribution of
electron-positron pairs to \Lg, which becomes a non-monotonic
function of $B$ (see Figs. 7 and 8 of
\cite{dyks}).  One of the consequences of this
modification is that expected values of \Lg~ for millisecond
pulsars, including J0437-4715, are typically about one or two
orders of magnitude lower than for classical pulsars (see Fig.\ref{gamma} in section~\ref{Xgamma}).
Quantitatively similar estimate, though for a different reason,
comes from the model of 
\cite{ds94}.
If these estimates  are correct then some millisecond
pulsars, including J0437-4715 (see Fig.\ref{j0437}), should be detected with
next-generation experiments, providing thus a testing ground for
magnetospheric models relevant for millisecond pulsars. 
GLAST is expected to have sensitivity above $100\MeV$ about 30
times higher then EGRET, and it will reach high-energy limit at
$300\GeV$ - closing thus for the first time the energy gap
between the VHE (very high energy) domain accessible with
ground-based Cherenkov techniques and the HE (high energy) domain
of satellite experiments \citep{glast}.  Moreover, 
millisecond pulsars are expected to radiate mostly in the range between $10\GeV$ and $100\GeV$, this
class of objects should be of interest for the upcoming IAC
telescope MAGIC \citep{blanch}.

For the time being, the only millisecond pulsar marginally detected above 100 MeV (at $3.5\,\sigma$ level) 
is PSR J0218$+$4232,
with the spin period $P = 2.3$ ms and the inferred magnetic field 
$B_{\rm pc} \simeq 8.6\times 10^8$ G, \citep{vkb1996}.
The inferred luminosity of the pulsed emission for 1 steradian opening angle
reaches $L_\gamma \simeq 1.64\times 10^{34}\ergs \simeq 0.07\, L_{\rm sd}$.
Broad-band  spectrum of this 
pulsar 
differs from the HE spectra known for young gamma-ray 
pulsars:
above 100 MeV the photon index $\alpha_{\rm ph} \sim -2.6$ \citep{kuiper_milli}  
and the spectrum 
resembles soft spectra of EGRET UID sources;
within the BeppoSAX range the spectrum is extremely hard, with  
$\alpha_{\rm ph} \simeq -0.61\pm 0.32$ \citep{mck2000}.
\begin{figure}
\vspace*{0.6cm}
\centerline{\psfig{file=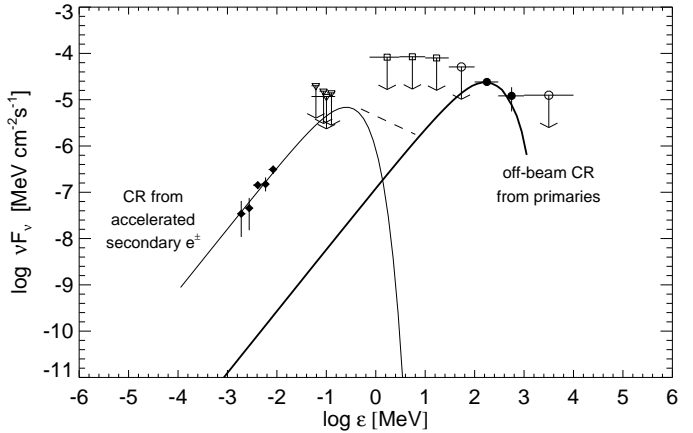,width=9.2cm} }
\caption[]{
Two-component theoretical  spectrum for the off-beam geometry
(continuous lines)  is compared with the spectrum observed for PSR J0218+4232
(diamonds -- BeppoSAX, triangles
-- OSSE, squares -- COMPTEL, circles -- EGRET). 
The low energy component 
is due to curvature
radiation from  secondary $e^\pm$ pairs
(with energy $\sim 10^5$MeV acquired upon their acceleration in a residual electric field), 
whereas the high-energy component
is due to curvature
radiation from  primary electrons (with energy $\sim 10^7$MeV). (From \cite{dr2002}.)
}
\label{j0218}
\end{figure}
Off-beam 
geometry (i.e. the line of sight misses the hollow cone beam)
was recently proposed \citep{dr2002} for this pulsar.
In such a situation the high energy ``cutoff" occurs at a relatively low 
energy $\sim 100$ MeV (marked
by two EGRET points, see Fig.\ref{j0218}). Unlike in on-beam cases, 
this cutoff is not due to magnetic absorption; it corresponds to
a maximum energy of those electrons which emit curvature photons towards the observer. 
The shape of the apparent cut-off can easily 
mimic very soft spectra
(\cite{hz2001} used similar viewing geometry arguments
to explain very soft spectra of  EGRET UID sources).
The BeppoSAX points are explained in this model 
as due to
curvature radiation of secondary $e^\pm$ pairs accelerated 
within the polar gap. 
Upon the acceleration in a residual electric field, 
the $e^\pm$ pairs should reach the energy $\gamma_\parallel$  between $1.5\times10^5$ and
$6\times 10^5$, which is about $1$\% of the primary electron energy. 
A few hundred of pairs per primary electron are required to be accelerated in order
to reproduce the BeppoSAX level. 
According to  \cite{hm2001} and~2002,
acceleration of such a large number of pairs within the polar gap is difficult to achieve.

\section{Viewing geometry effects}
Viewing geometry has been recognized as an important factor in shaping the
lightcurves as well as broadband energy spectra  of 
particular objects (like e.g. the Vela pulsar, see \cite{hm98})
as well as
in the context of EGRET UID galactic sources (e.g. recent review by \cite{b2001}).

\begin{figure}
\begin{center}
\includegraphics[width=0.5\textwidth]{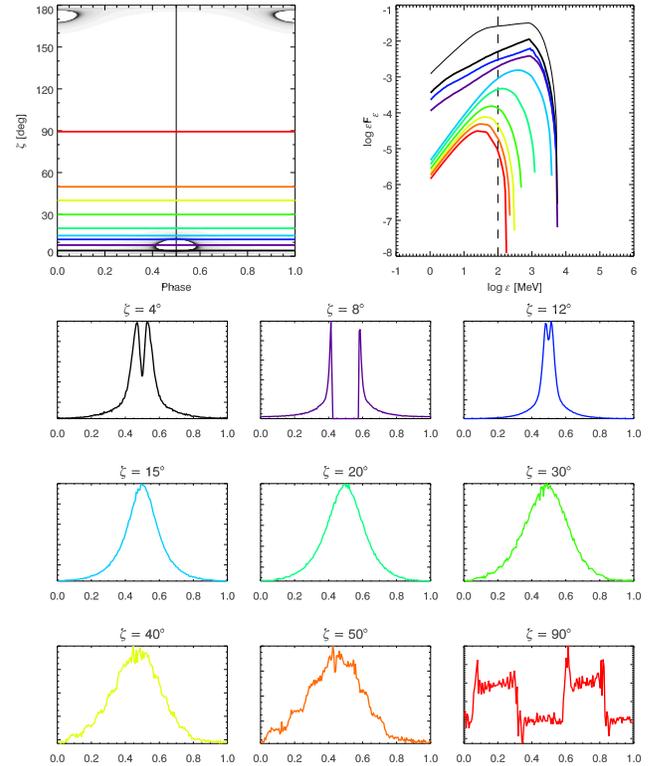}
\end{center} 
\caption[]{The characteristics of high energy radiation expected from a classical
pulsar with $P=0.1$\,s, $B=10^{12}$\,G and the magnetic dipole inclination  
$\alpha = 8^{\circ}$. The top left panel shows the intensity distribution
for radiation above $100$\,MeV as a function of rotational phase~$\phi$ 
and viewing angle $\zeta$. Horizontal color-coded lines  correspond to the
position of 
nine observers located at different angles $\zeta$. 
For each $\zeta$ the phase-averaged
spectrum drawn
with  color-coded line
is shown in top right panel.
The corresponding pulse profiles  
are shown in small panels labelled with $\zeta$.
The spectrum of total luminosity from the pulsar (thin
solid line in top right panel) is plotted (at an arbitrary level relative to the 
phase-averaged spectra) for comparison of the shapes. (From \cite{wozna}).
}
\label{ctilt8}
\end{figure}
  
\begin{figure}
\begin{center}
\includegraphics[width=0.5\textwidth]{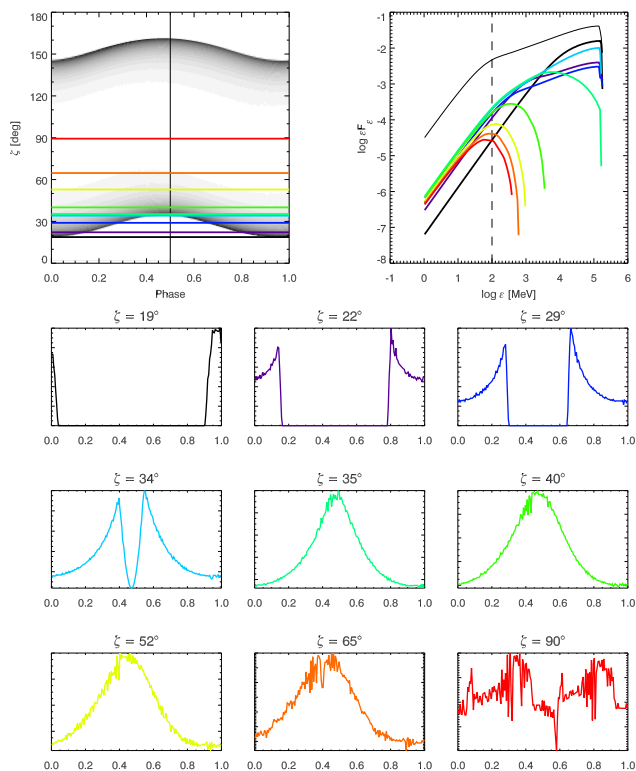}
\end{center} 
\caption[]{Same as fig.\ref{ctilt8}, but for a millisecond pulsar with
$P=2.3$\,ms, $B=10^{9}$\,G and different set of viewing angles $\zeta$. (From \cite{wozna}).
}
\label{mtilt8}
\end{figure}

To taste a variety of apparent pulsar characteristics accessible
upon changing the viewing angle $\zeta$ we present here two cases: of a classical pulsar
in Fig.\ref{ctilt8} and a millisecond pulsar in Fig.\ref{mtilt8}. 
The lightcurves (these are energy-integrated above $100\,$MeV) and the gamma-ray spectra 
were calculated with our numerical code.
The spectra are two-component: due to curvature and synchrotron emission; inverse Compton scattering was neglected.
The effects of aberration and light travel delays 
were included. These effects are important for millisecond pulsars, and affect their
lightcurves, especially at large inclination angles $\alpha$ (the case not shown here).
Whenever the line of sight misses outer rim of the polar cap, the observed spectra
become soft with an apparent exponential (and not super-exponential) decay,
mimicking thus an outer-gap candidate. The millisecond gamma-ray pulsar J0218+4232 (Fig.\ref{j0218})
is probably a good example of such a case.
  
\section{Concluding remarks}
High-energy astrophysics of pulsars was challenged by
unexpected richness of spectral and temporal properties 
found for the brightest gamma-ray pulsars.
Numerous modifications (both, minor and major) to the existing models of magnetospheric activity
are being invented to accommodate at least some of these properties. 
It is clear, however, that
we actually need  good quality high-energy data for much weaker sources.
Only then it will be possible to asses on statistical grounds
the significance of those properties (the argument used by many authors, 
recently reiterated strongly by \cite{b2001}). Chances are that it may have happen
in a few years time.

The planned observatory GLAST \citep{glast}
will be superior to  EGRET  in two
aspects. First, its sensitivity at 10~GeV will be more than two orders of magnitude better than
that of EGRET. Second, it will reach energy of 300~GeV, closing thus for the first time
the energy gap between satelite and ground-based  observatories (HE and VHE, respectively).
The MAGIC Telescope \citep{blanch} -- a 17 m diameter Imaging Air Cherenkov Telescope  (IACT) -- is expected to 
operate with sensitivity about three orders of magnitude higher at 10~GeV
than EGRET. Its advanced technology will make 
possible to cover energy range  between 10~GeV and 1~TeV, and to reach $\sim 50\,$TeV in the
Large Zenith Angle mode. 
Energy ranges of GLAST and MAGIC will overlap over more than one decade in energy. 

We expect that the question of whether
millisecond pulsars emit gamma-rays at all, will be answered relatively easily with these 
high-sensitivity observatories, 
verifying at first the status of  J0218$+$4232.\\
In the context of numerous positive detections  anticipated at hard gamma-rays,
inclusion of viewing geometry effects will be of particular importance to properly interpret
the observed shapes of spectral turnovers at high-energy ends.

To tackle high-quality data of the future in the most effective way,
numerical 3D codes should be developed, capable of
fast tracing the development and propagation of
electromagnetic cascades in variety
of non-dipolar realizations of magnetic field structure.
Relaxing the centered-dipole assumption as a first step  would be very much in line with
recent conclusions from the
soft X-ray data analysis for neutron stars (Pavlov et~al.~\cite{pavlov2002}).
Optical to gamma-ray properties of Vela (section~\ref{problems}) are in our opinion a signature of
a strongly non-axisymmetric 
hollow cone of radiation around the magnetic axis. Strong deviations of the actual magnetic field structure
from a pure dipole at the stellar surface 
may well be responsible for inducing
this axial asymmetry already at the site of electron acceleration.
(Recently, multipolar character of open field lines has been considered in a quantitative way by Gil et~al.~\cite{gil}
to justify radiopulsars with vacuum gap solutions in pulsars with superstrong magnetic fields.)
An advantage of incorporating  multipolar components to the polar-cap models is twofold: \\
1) magnetic field strength may locally
be lowered slightly, opening thus windows for GeV photons to escape without magnetic attenuation;
postulating high-altitude accelerators would be then unnecessary.\\
2) directional characteristics of high-energy radiation would change notably with respect
to the dipolar axis at higher altitudes, where radio emission is generated.

\begin{acknowledgements}
BR acknowledges financial support by Wilhelm und Else Heraeus-Stiftung; he also 
wishes to thank the organizers of the Bad-Honnef Seminar for invitation and warm hospitality.
This work was supported by the KBN grants: 2P03D02117 (BR, TB) and 5P03D02420 (JD).

\end{acknowledgements}

\end{document}